\documentclass[sigconf]{acmart}

\usepackage{booktabs} 

\setcopyright{rightsretained}

\usepackage{amssymb}
\usepackage{graphicx}
\usepackage{subcaption}
\usepackage{amsmath,amssymb}
\usepackage{float}
\usepackage{array}
\usepackage{algorithm} 
\usepackage{tabularx}
\usepackage{enumitem}

\usepackage{algorithm} 
\usepackage[noend]{algpseudocode} 

\usepackage{amsmath}               
           {
             
           }
\graphicspath {{figures/}}
\DeclareMathOperator{\E}{\mathbb{E}}

\acmDOI{}

\acmISBN{}

\acmConference[]{}{}{} 
\acmYear{2018}
\copyrightyear{2018}

\acmPrice{15.00}

\begin{document}
\title[]{Semisupervised Learning on Heterogeneous Graphs and its Applications to Facebook News Feed}

\author{Cheng Ju}
\affiliation{%
  \institution{Facebook, Inc\\ University of California, Berkeley}
}

\author{James Li}
\affiliation{%
  \institution{Facebook, Inc}
}

\author{Bram Wasti}
\affiliation{%
  \institution{Facebook, Inc}
}

\author{Shengbo Guo}
\affiliation{%
  \institution{Facebook, Inc}
}

\renewcommand{\shortauthors}{}

\begin{abstract}

Graph-based semi-supervised learning is a fundamental machine learning problem, and has been well studied. Most studies focus on homogeneous networks (e.g. citation network, friend network). In the present paper, we propose the Heterogeneous Embedding Label Propagation (HELP) algorithm, a graph-based semi-supervised deep learning algorithm, for graphs that are characterized by heterogeneous node types. Empirically, we demonstrate the effectiveness of this method in domain classification tasks with Facebook user-domain interaction graph, and compare the performance of the proposed HELP algorithm with the state of the art algorithms. We show that the HELP algorithm improves the predictive performance across multiple tasks, together with semantically meaningful embedding that are discriminative for downstream classification or regression tasks.
  
\end{abstract}

%




\maketitle

\section{Introduction}

\subsection{Motivation}

There are multiple factors that influence the ranking of a story on a person's News Feed. A comprehensive look of the many factors involved can be found in \citep{newsfeedlars}. For content that contains links to outside Web domains, one of the most important factors is the quality of the content from this domain. There are different dimensions under consideration for the overall quality of a domain (e.g. if its URLs always contain exaggerated headlines). For many of the important dimensions, we train classifiers to predict the likelihood a piece of content is of this dimension using content features. These classifier predictions are then used in conjunction with other signals (e.g. timeliness, interaction history) to assess the content rank on a person's News Feed.

We have following demands and expectations for the semi-supervised methods for our applications. First, as the data is large and predictions can get stale quickly, we must pay special attention to training time and warm-start issues. When an unseen domain appears, we need the score immediately, instead of retrain the model on the whole data. Second, as the number of nodes is huge, if the embedding is given by a look-up table for every nodes in the graph, the computation would be a bottleneck. Thus we plan to avoid embedding nodes based on a look-up table. Third, as we has clear classification tasks, we are looking for an end-to-end approach to take the graph information into supervised training simultaneously, instead of two-stage embedding-supervision procedures. Last but not least, the quality labels $y$ are usually obtained by human labeler and thus expensive, while the content features $X$ and different type of interaction graphs (e.g. resharing graph) are easy to get. The inductive semi-supervised methods have great advantage under this condition.

\subsection{Graph-based Semi-supervised Learning}

Graph-based semi-supervised learning  is widely used in network analysis, for prediction/clustering tasks over nodes and edges. A class of commonly used approaches can be considered as a two-stage procedure: the first first step is node embedding, where each node are represented in a vector which contains the graph information; the second step simply apply these vectors are further for the conventional machine learning tasks.
\cite{ng2002spectral} proposed a spectral clustering method, which uses the  eigenvectors of the normalized Laplacian matrix as node embedding, and applies k-means algorithm on the embedding vectors for unsupervised clustering.
\cite{newman2006finding} proposed another clustering method, using the eigenvectors  of the modularity matrix to find hidden community in networks .
\cite{al2006link} generated several  handcrafted local features (e.g. sum of neighbors) as embedding, and applied supervised learning  on them to predict the probability that two node would be connected in the future, which is more flexible compared to proximity based link-prediction \cite{liben2007link,lu2011link}.
\cite{tang2009relational,tang2011leveraging} further studied the embedding methods proposed by  \cite{ng2002spectral,newman2006finding} for supervised learning tasks, to predict the community label of the nodes in social network, which showed great success.
\cite{tang2009scalable} proposed a edge-centric clustering scheme, which learns
a sparse social dimension for each node by clustering its edges. Recently,
several deep learning based representation learning methods have shown great
success in a wide range of tasks for network data. DEEPWALK
\cite{perozzi2014deepwalk} learns latent representations of vertices in a
network based on truncated random walks and the SkipGram model. Node2vec
\cite{grover2016node2vec} further extends DEEPWALK by two additional bias search
parameters which controls the random walks, and thus control the representation
on homophilic and structural pattern. Both of \cite{perozzi2014deepwalk} and
\cite{grover2016node2vec} are assessed by feeding the generated embedding into a
supervised task on graph. Compared to previous embedding methods, these two
methods are more flexible and scalable: the features could be learned by
parallel training with stochastic gradient descent, and adding new nodes on the
graph does not require recomputing the features for all the observations. With
extra computational trick like negative sampling and hierarchical loss
\cite{mikolov2013distributed}, the computation could be further reduced. To
learn sparse features, \cite{chang2015heterogeneous} further proposed a deep
learning based model for the latent representation learning of mixed categories
of vertex. Large-scale information network embedding \cite{tang2015line} computes the embedding by optimizing the objective function to preserve  ``first-order'' and ``second-order'' graph proximity.

Another class of semi-supervised methods directly use the graph information during supervised training, instead of the two-stage embedding-supervision procedure in the last paragraph. Label propagation \cite{zhu2002learning} is an simple but effective algorithm, where the label information of labeled nodes are propagated on graph to unlabeled data.
  \cite{yang2016revisiting} presented a semi-supervised learning framework that learns graph embedding during the training of a supervised task.  \cite{yang2016revisiting} further proposed both transductive and inductive version of their algorithm, and compared them with several widely used semi-supervised methods. 
  The neural graph machine  \cite{bui2017neural} extended idea of label propagation of regularizing on the final prediction to regularizing the hidden output of neural networks.
  Another class of algorithms build additional nuisance task to predict the graph context, in addition to the supervised label prediction.

  Most work about semi-supervised learning on graph focused on homogeneous networks, where there exists only singular type of nodes and relationships. LSHM (Latent Space Heterogeneous Model) is proposed by \cite{jacob2014learning}, which creates a look-up table for the embedding of each node in the graph. The model are trained by both the supervised loss, defined as classification loss from a logistic regression model on the top of the embedding, and an unsupervised loss, defined as the distance between two connected nodes. \cite{chang2015heterogeneous} further proposed the Heterogeneous Networks Embedding (HNE) algorithm based on deep neural networks, which in contrast is a purely unsupervised method. It uses each pair of node  as input to predict their similarity, and define a hidden output as the embedding. It applies different network structure to process  nodes with different type, while keeps the networks sharing the parameter for same type of node. Inspired by DeepWalk and Node2vec, \cite{dong2017metapath2vec} proposed a new meta-path-based random-walk strategy to build the sequences of nodes, and then feed them into SkipGram model to get a unsupervised embedding for each node.

  In this work, we propose a new graph-based semi-supervised algorithm, HELP (heterogeneous embedding label propagation). It is an inductive algorithm that can utilize both the features and the graph where predictions can be made on instances unobserved in the graph seen at training time. It is also able to handle multiple heterogeneous nodes in the graph, and generate embeddings for them. We call it ``label propagation'' as it also implicitly impose a ``smooth constraint'' based on the graph \cite{bui2017neural}, which is similar to the label propagation algorithm \cite{zhu2002learning}. We also demonstrated the effectiveness of our proposed approach with several node-classification tasks on a subset of the Facebook graph consisting of users and Web Domains, with focus in particular to identifying domains who repeatedly show content that are sensational \cite{feedfyicb} and/or otherwise low quality \cite{feedfyiaf}, or domains who repeatedly show content that are authentic and high quality \cite{feedfyihq}.

\subsection{Notations}

We use the notation $u_i$ to denote the feature vector for the $i$-th user.  We use $d_j$ to denote the feature vector for the $j$-th domain. We use $y_j$ to denote the label of domain $d_j$. We use the index $j = 1, \cdots, L$ to denote the index of the labeled domains. We further define a function $\text{concat}(\cdot, \cdot)$, which concatenates two row vectors into one. We use $X^T$ to denote the transpose of a matrix $X$. We use $\theta$ to denote all the trainable model parameters for a neural network.

\subsection{Related Works}

\begin{figure}[h]
  \centering
  \includegraphics[width=0.35\textwidth]{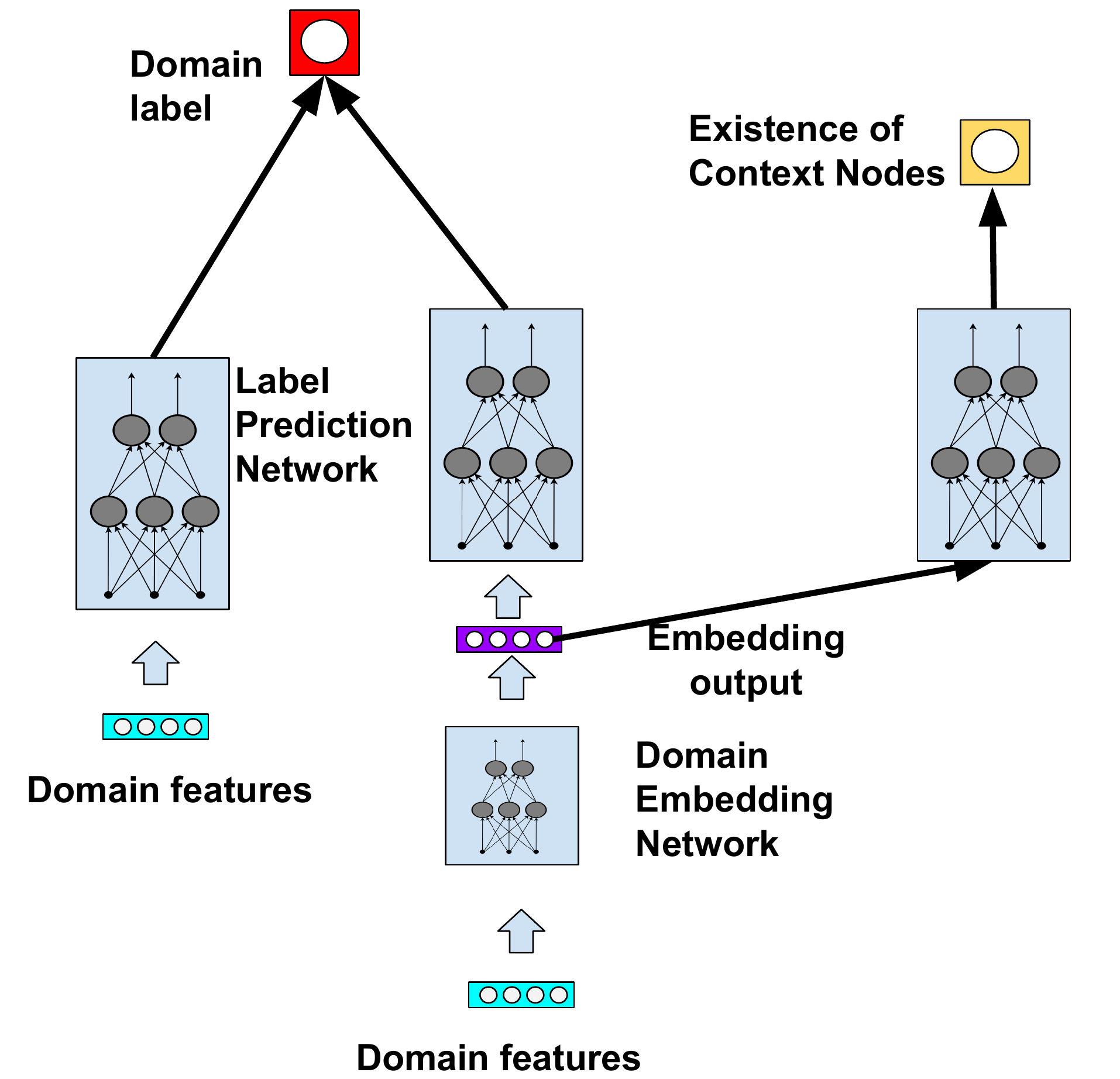}
  \caption{Network architecture for Planetoid-I.}
  \label{fig:planetoid}
\end{figure}

In this section, we review several inductive contextual graph-based semi-supervised deep learning methods, and show how they can be applied into our domain classification task. In general, graph-based semi-supervised learning  methods relies on the assumption that connected nodes tend to have similar labels. By this assumption, \cite{yang2016revisiting} summarized that the loss function for graph-based semi-supervised learning can be decomposed into two part: the supervised loss part (fitting constraint) and the graph-based unsupervised regularization part (smoothness constraint).
\cite{yang2016revisiting} systematically summarized most of the non-deep graph-based semi-supervised learning method, including Learning with local and global consistency \cite{zhou2004learning} and Manifold regularization \cite{belkin2006manifold}. It then presented a semi-supervised learning framework called Planetoid (Predicting Labels And Neighbors with Embeddings Transductively Or Inductively from Data) that learns graph embedding during the training of a supervised task. Authors further proposed both the transductive and inductive version of their algorithm, and compared them with several widely used semi-supervised methods \cite{yang2016revisiting}. Figure \ref{fig:planetoid} shows the inductive version of the Planetoid with an example on our domain label prediction task, where the features are passed into a feed-forward neural network for both predict the domain label and the graph context. The transductive version is similar, except it trains a look-up table for each domain as embedding, instead of the intermediate output of a neural network (a parameterized function of input feature vectors). In this example, the supervised loss is the label classification loss for each domain, and the unsupervised loss is defined as the prediction loss for the existence of each domain in its context, where the context is defined for the nodes share the same label, or the nodes appear close to each other in the random walk on the graph based on DEEPWALK \cite{perozzi2014deepwalk}.

To be more specific, the right-most network block in figure \ref{fig:planetoid} used in \cite{yang2016revisiting} is a single-layer network with sigmoid activation and $w_c$ is the row for node $c$ in the context matrix \cite{goldberg2014word2vec}, which makes the loss function for Planetoid-I to be:

\begin{eqnarray*}
 G_{Planetoid-I}(\theta)&=& L_s + L_u\\
L_s &=&   - \frac{1}{L} \sum_{i=1}^{L} \log p(y_i|d_i) \\
 L_u &=& \lambda \E_{i, c, \gamma} \log \sigma(\gamma w_c^T h(d_i))
\end{eqnarray*}
where $\gamma$ is a binary indicator determines if the nodes with index $c, i$ are similar or not; $p(y_i|d_i)$ is the final output on top of the left three building blocks, representing the predicted probability of true label from the classification neural network. $h$ represents the building block at the middle bottom, which generates the embedding for the node by applying a parametric function on the input feature vector. $\lambda$ is the hyper-parameter that controls the trade-off for the fitting constraint and smoothing constraint.

The neural graph machine  \cite{bui2017neural} is a deep learning based extension of label propagation, which imposes a non-linear smoothing constraint by regularizing the intermediate output of a hidden layer of neural networks. In our example, the supervised loss is still the predicting loss for the domain label, while the unsupervised smooth constraint is the average distance between connected domains. 


\begin{eqnarray*}
 G_{NGM}(\theta)&=& L_s + L_u\\
L_s &=&   - \frac{1}{L} \sum_{i=1}^{L} \log p(y_i|d_i) \\
L_u &=& \lambda_1 \sum_{i, j \in \mathcal{E}_{LL}} w_{d_i, d_j} d(h(d_i), h(d_j)) + \\
&&\lambda_2 \sum_{i, j \in \mathcal{E}_{LU}} w_{d_i, d_j} d(h(d_i), h(d_j)) \\
&& \lambda_3 \sum_{i, j \in \mathcal{E}_{UU}} w_{d_i, d_j} d(h(d_i), h(d_j)) 
\end{eqnarray*}
where $d(\cdot, \cdot)$ is a distance function for a pair of vectors, and \cite{bui2017neural} suggests define $d$ with $l1$ or $l2$ distance for the two input vectors. $p(y_i|d_i)$ has same meaning as for Planetoid-I, and $h(d_i)$ is the node embedding that defined as the intermediate output of the second laster layer. $\mathcal{E}_{LL}$, $\mathcal{E}_{LU}$ and $\mathcal{E}_{UU}$ defines the node pair that both labeled, only one labeled, and both unlabeled. $\lambda_1,\lambda_2, \lambda_3$ are hyper-parameters control the smoothing constraint for different label types.

\begin{figure}[h]
  \centering
  \includegraphics[width=0.35\textwidth]{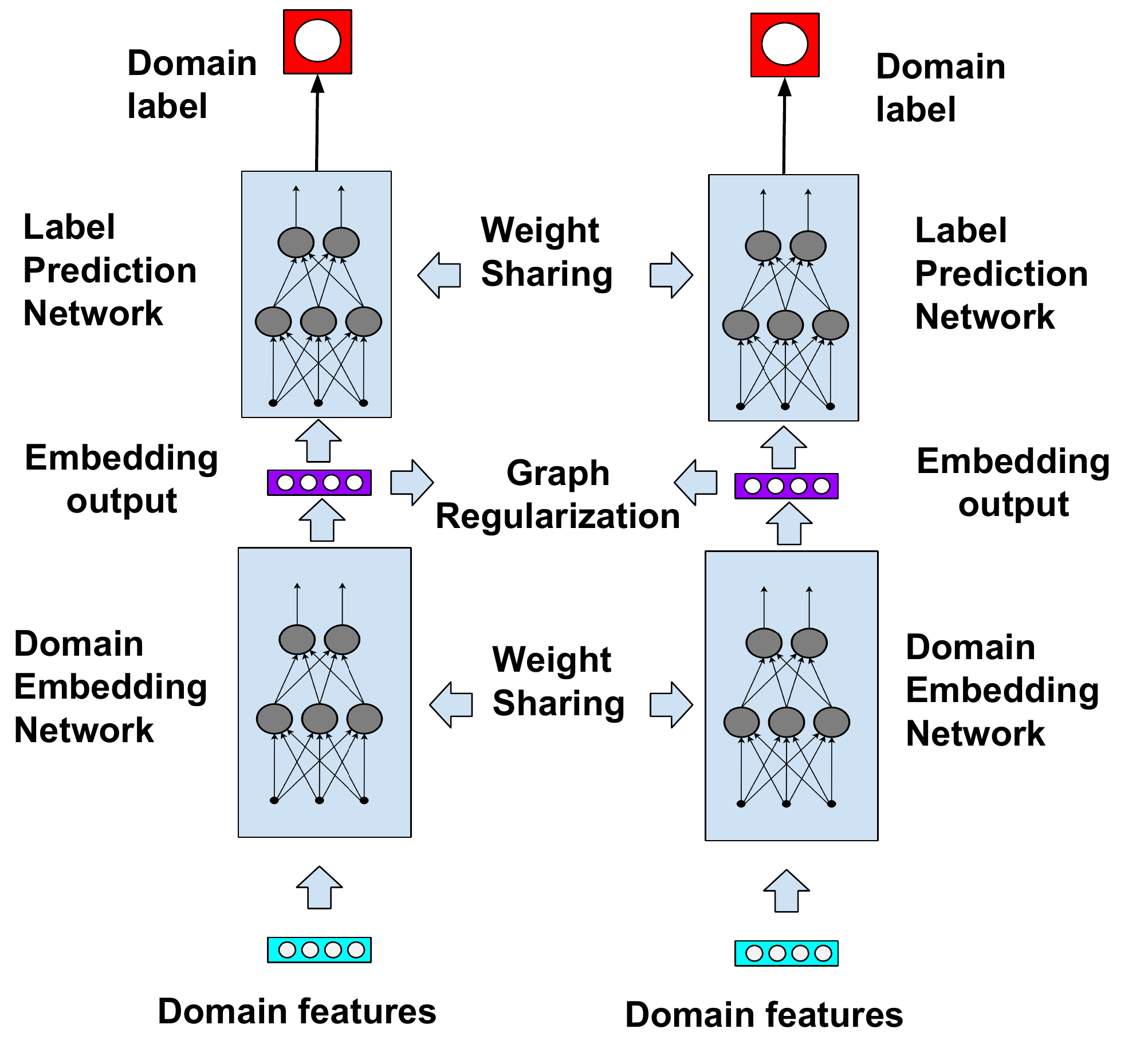}
  \caption{Network architecture for neural graphical machines}
  \label{fig:ngmh}
\end{figure}
  
\section{The HELP}

\subsection{Neural Network Structure}

\begin{figure}[h]
  \centering
  \includegraphics[width=0.35\textwidth]{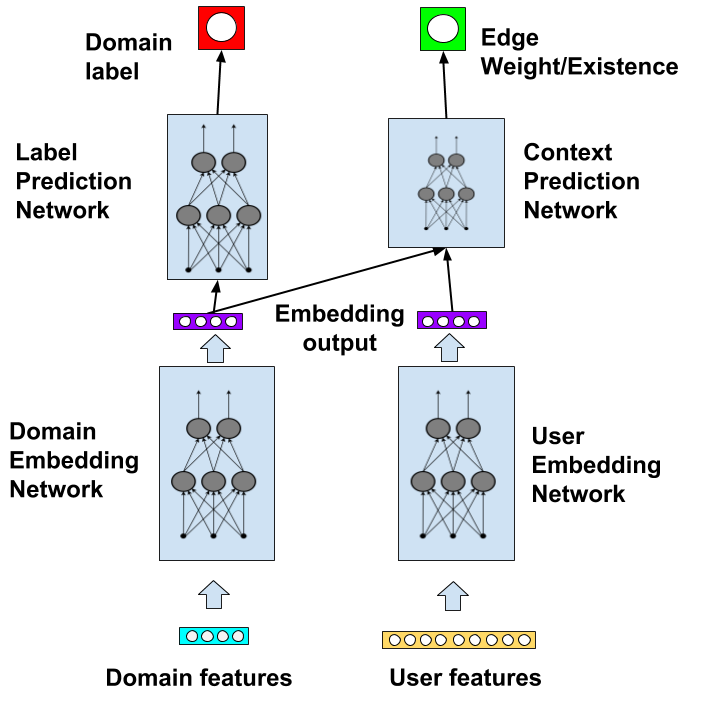}
  \caption{The network structure for the HELP.}
  \label{fig:help}
\end{figure}

Figure \ref{fig:help} shows the network structure of the HELP for user-domain network, where the nodes in the network are either domain or user, and the links are defined as the interaction between users and domains (e.g. an user likes a domain). Inspired by the neural graphical machines \cite{bui2017neural}, which impose a smoothing constraint on the intermediate output of a feed forward neural network, we propose a new network architecture with four building blocks that can handle two different type nodes. The two building blocks, $h_d, h_u$, at the bottom of figure \ref{fig:help} represents two feed forward neural network block, with the input as the contextual features of domain and user, and the output as the embedding for domain and user. Two ``embedding'' building blocks do not share any parameter, and there is no constraint on the input/output shape.

After the two-tower  ``embedding'' building blocks, we define the other two building blocks. The first is the label prediction block for domain label prediction, which we defined as $f$. It takes the embedding $e_d = h_d(d_i)$ of the given domain as input, and output the probability $f(e_d)$ that the given domain would be labeled 1 by human checker. The other is the ``context'' block $g$, which ``predicts'' the context of the graph. To be more specific, it is a block of feed forward neural network that computes the distance $g(e_u, e_u)$ of between the user and the domain, given the embedding of both of them from the embedding blocks.

During the training stage, the inputs are the pairs of the user-domain. Inspired by \cite{bui2017neural}, our proposed objective function function can be also decomposed into a neural network cost (supervised) and the label propagation cost (unsupervised) as follows:

\begin{eqnarray*}
  G_{HELP}(\theta) &=& \sum_{j=1}^{L}L_{s}(f(h_d(d_j)); \theta ) +\\
  && \lambda \sum_{i, j} L_{u}(w_{u_i, d_j}, h_d(d_i), h_u(u_i))
\end{eqnarray*}

The first part, the supervised loss, is the cross-entropy for the binary label of domains:

$$L_{s}(f(h_d(d_j)) = y_{j}\log(f(h_d(d_j)) + y_{j}\log(1-f(h_d(d_j))$$

The second part, the graph regularization loss, is defined as:

\begin{eqnarray*}
  L_{u}(w_{u_i, d_j}, h_d(d_i), h_u(u_i)) &=& w_{u_i, d_j} \cdot d_{u_i, d_j}^2\\
  &&+  (1 - w_{u_i, d_j})  \cdot \max(0,m - d_{u_i, d_j})^2
\end{eqnarray*}
where $ d_{u_i, d_j} =\sqrt{1- g(\text{concat}(h_d(d_i), h_u(u_i))})$, and $m$ is a tunable, fixed margin parameter. Having a margin indicates that unconnected pairs that have the distance beyond this margin will not contribute to the loss. This loss is used in Siamese network, to distinguish a given pair of images \cite{koch2015siamese}. Instead of using L2 distance of the output of an embedding network/feature extractor, we use a separate neural network block to generate ``similarity score'' for each pair, and use one minus such score as the distance metric. 

In our experiment, the input contextual features are numerical, thus we only consider the fully-connected neural networks. We can  easily extend it with image features using convolutional neural networks, and text features using recurrent neural networks.  $f$ is a 2-layer fully connected neural network with output shape $(16, 1)$; $h_d$ and $h_u$ are both 3-layer fully connected neural networks with output shape $(96, 64, 32)$, without parameter-sharing; $g$ is a 2-layer fully connected neural network with output shape $(16, 1)$.

During the training phase, in each epoch, all the labeled domain are passed, and user-domain pairs are sub-sampled due to the huge number of pairs. In each iteration within the epoch, the total loss is computed, and the gradient based on the total loss is back-propagated to the whole network, including $f$, $g$, $h_u$, and $h_d$, simultaneously. During the domain classification/predicting stage,  it requires no extra re-training, and only the domain feature is used.

Notice here the network structure is designed for user-domain bipartite graph. It can be  adapted to multiple type of nodes, with multiple smoothing constraints for more than one edge type.

\section{Experiments}

\subsection{Labels of Domains}

The labels used in the experiments are generated manually according to some internal guideline. We consider three different ``dimensions'': each dimension stands for a certain type of quality evaluation. Table \ref{tab:stat} shows the summary statistics of each label.

\begin{table}[ht]
  \caption{Summary Statistics for Labeled Domains}
  \label{tab:stat}
  \begin{center}
    \begin{tabular}{c|c|c| }
      Label Type & Total Size & \# of Positive  \\ \hline
      dimension\#1 & 5498 & 1094\\
      dimension\#2 & 6399 & 748\\
      dimension\#3 & 1781 & 477\\
      \hline
    \end{tabular}
  \end{center}
\end{table}

\subsection{Metric}
\label{sec:metric}

In the experiments, we considered a binary classification problem, thus following metrics are considered.
The first metric is the area under Receiver Operating Characteristic curve (AUROC). The curve is plotted with the true positive rate (TPR) against the false positive rate (FPR) at various threshold settings. The AUROC is defined as the area below the ROC curve. It can be explained as the expectation that a uniformly drawn random positive is ranked before a uniformly drawn random negative.

The second metric is the area under the Precision-Recall curve (AUPRC). The curve is plotted with the precision (true positives over the sum of true positives and false positives) against the recall (true positives over the sum of true positives and false negatives) at various threshold settings. In practice, we are more in favor of AUPRC in comparison to AUPRC due to the following reasons. First, the classes for all the three label types are imbalanced. It has been shown that in the imbalanced data set, PR curve is more informative \cite{saito2015precision}. To be more specific, as there are much more negative samples than positive ones, the true negative examples will overwhelm the comparison in ROC, while will not influence PRC. The second reason is we mainly focus on finding the positive (the domains labeled as 1). The AUPRC mainly reflect the quality of retrieval of the positives and its value is not invariant when we change the baseline (which category should be labeled as 1), which is different from AUROC.



\subsection{Features}

For domains, we collected 29 content features, which include multiple summary statistics (e.g.  number of likes), and some ranking score generated from other model. For users, we collected 129 background features, most of which are user activity statistics in the past. We do not disclose the details of features as it does not influence understanding the proposed algorithm and the following experiments.

We sub-sampled $2.4$ million English-speaking users at Facebook for this offline experiment, with the domains that  have at least one interaction with the sampled users in last 7 days. The bipartite graph contains $14.46$ Million user-domain edges.

\begin{table}[ht]
\begin{center}
  \caption{Sample size for user, domain, and their interactions (edges).}
  \begin{tabular}{c|c }
    type & size \\ \hline
    Domain &  $241,205$ \\
    User &  $2,433,581$ \\
    Edge & $14,460,336$ \\
    \hline
  \end{tabular}
\end{center}
\end{table}

\subsection{The User-Domain Graph}

\begin{figure}[h]
  \centering
  \includegraphics[width=0.25\textwidth]{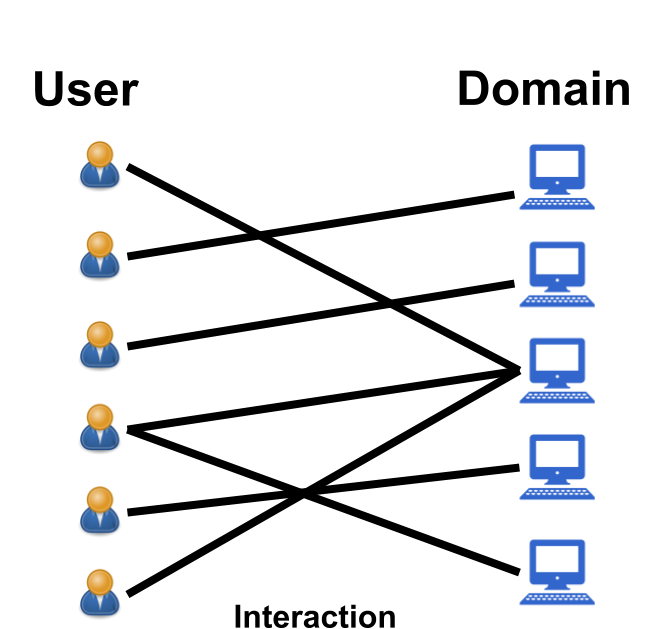}
  \caption{An illustration of user-domain interaction graph.}
  \label{fig:user_domain_graph}
\end{figure}

Figure \ref{fig:user_domain_graph} visualize a user-domain graph. 
Each edge is considered as undirected, containing two information: the interaction type, and the count of such interaction in last 7 days. In this study, we only focus on the Resharing as the interaction type. Thus the weight of each edge represents the number of reshares for the given user for the URLs from the given domain.

The experimental data is generated on 10/27/2017, which means the graph is based on the user-domain interaction statistics from 10/20/2017 to 10/27/2017.

\section{Benchmarks}

We consider following algorithms as benchmarks for HELP:
\begin{itemize}
\item Label Propagation algorithm (LP) by \cite{zhu2002learning}, which only use the graph information. It is not surprising to see it has much worse performance compared other methods use the more informative contextual features. We report this only to demonstrate much information contains in the graph.
\item Multi-layer Perceptron (MLP), which is a fully connected feed-forward neural network  using only the  feature information.
\item Planetoid-I (Predicting Labels And Neighbors with Embeddings Transductively Or Inductively from Data, Inductive Version) by \cite{yang2016revisiting}, with domain-domain graph compressed from user-domain graph.
\item Neural Graph Machine (NGM) by \cite{bui2017neural}, with domain-domain graph compressed from user-domain graph.

\end{itemize}

As we don't have explicit domain-domain graph, we construct it by compressing the user-domain graph. we construct the domain-domain graph by:

\begin{enumerate}

\item For each domain $d_i$ and domain $d_j$, find the set of users $U$ have edges for both domains.

\item For each user $u_k \in U$, define the similarity of two domains based on the similarity between user domains:  $sim^{d_i, d_j}_k = \min(e_{u_k, d_i}, e_{u_k, d_j})$.

\item Finally define the edge between $d_i, d_j$ as the sum of the similarities computed from all users: $$e_{d_i, d_j} = \sum_{u_k \in U} sim^{d_i, d_j}_k.$$

\end{enumerate}

There are multiple way to compress the user-domain graph to domain-domain graph. We have experimented multiple strategies, but does not show significant difference. As this is not the main focus of this study, we only choose the most straightforward one.

\subsection{Optimization}

All the neural network models are trained by Adam optimizer \cite{kingma2014adam}, with initial learning rate $0.001$, and decayed with ratio $0.1$ for every 20 epochs. We set the weight decay as $10^{-5}$. We train each model 60 epochs. We train each network 10 times and report the average of each performance metric as this can stabilize the results by reducing the impact of randomness in initialization and training \cite{ju2017relative}. 

We also experimented warm-start reported in \cite{yang2016revisiting}. However, this does not improve the performance. So the supervised and unsupervised part are trained simultaneously. 

\section{Classification Performance}

\subsection{Experiment Results}

Though two metric are reported, we mainly focus on the AUPRC, as we mainly want to improve the quality of retrieval for positive samples. See  detailed discuss ion section \ref{sec:metric}.

\begin{table}[ht]
  \caption{The predictive performance on testing set for dimension\#1 domain label. All the values are in $10^{-2}$ scale.}
  \label{tab:labeled-dimension1}
\begin{center}
  \begin{tabular}{c|c|c
    }
    Model & AUROC & AUPRC 
    \\ \hline
    LP & 85.7 & 71.0 
    \\\hline\hline
      MLP & 95.1  &  83.3  
      \\\hline\hline
    PLANETOID-I & 95.1 & 83.8 
    \\
    NGM L1 & \textbf{95.3}  &  83.5  
    \\
    NGM L2&  95.1 &  82.9 
    \\
    HELP & 95.2 & \textbf{84.2}  
    \\
    \hline
  \end{tabular}
\end{center}
\end{table}

Table \ref{tab:labeled-dimension1} shows the predictive performance when predicting if a domain should be labeled as a dimension\#1 domain. The AUCROC does not have noticeable difference for all deep learning based algorithms. For AUPRC, Planetoid-I and NGM with L1 regularization slightly improved the performance, and HELP achieved the best performance.

\begin{table}[ht]
  \caption{The predictive performance on testing set for dimension\#2 domain label. All the values are in $10^{-2}$ scale.}
   \label{tab:labeled-dimension2}
\begin{center}
  \begin{tabular}{c|c|c
    }
    Model & AUROC & AUPRC 
    \\ \hline
     LP & 87.1 & 67.7 
     \\\hline\hline
    MLP & 95.6 & 81.6  
    \\\hline\hline
    PLANETOID-I & 95.6  & 81.9 
    \\
    NGM L1 & 95.6 & 80.9  
    \\
     NGM  L2&  95.7 & 81.5 
     \\
    HELP & \textbf{96.3} & \textbf{82.9} 
    \\
    \hline
  \end{tabular}
\end{center}
\end{table}

Table \ref{tab:labeled-dimension2} shows the predictive performance when predicting if a domain should be  labeled as a dimension\#2 domain. Similar to previous experiment, the AUCROC does not have noticeable difference, which may due to the severe imbalance of the positive/negative samples. For AUPRC, the HELP significantly improved the benchmark MLP by $1.3\%$ absolute increment.  The Planetoid-I have small improvement compared to MLP, while other semi-supervised method does not show any noticeable improvement.

\begin{table}[ht]
  \caption{The predictive performance on testing set for dimension\#3 domain label. All the values are in $10^{-2}$ scale.}
   \label{tab:labeled-dimension3}
\begin{center}
  \begin{tabular}{c|c|c
    }
    Model & AUROC & AUPRC 
    \\ \hline
    LP & 71.9 & 50.1 
    \\\hline\hline
    MLP & 82.2 & 58.1  
    \\\hline\hline
    PLANETOID-I & 82.2  & 60.2 
    \\
    NGM L1 &   82.6 & 63.3 
    \\
     NGM  L2&  82.2  & 62.9 
     \\
    HELP &\textbf{82.6} & \textbf{64.9} 
    \\
    \hline
  \end{tabular}
\end{center}
\end{table}

Table \ref{tab:labeled-dimension3} shows the predictive performance when predicting if a domain should be labeled as a dimension\#3 domain. Different from previous two labels, all the semi-supervised learning methods significantly improve the AUPRC, with at least $2\%$ absolute improvement. One of the most convincing reason is the dimension\#3 data is much smaller than dimension\#1/dimension\#2 dataset, which is usually considered as the case that in favor of the semi-supervised method than purely supervised methods. The HELP model achieved best performance for both AUROC ($0.4\%$ absolute improvement) and AUPRC ($6.8\%$ absolute improvement).

\subsection{Comparison of Unsupervised Graph-based Loss}

There are many loss functions can be applied for ``context prediction'' in the graph-based neural networks. In this section, we investigated the performance for different several variations of the HELP with different semi-supervised loss function.

\subsubsection{Weighted Graph}
\label{sec:weight}

Then we first consider commonly used  supervised loss functions for edge prediction as the graph regularization.

After generates the embedding for an user $e_{u}$ and a domain $e_{d}$, we concatenate two embedding into one:
$$e_{\text{concat}} = \text{concat}(e_{u}, e_{d})$$
and directly feed it into a feed-forward neural network $g$ to predict the edge for this user-domain pair:
$$\hat{w}_{u, d} = g(e_{\text{concat}})$$

In this setting, the label is the weight of the edge (i.e. number of reshares in the past week). We considered the following loss functions:

\begin{itemize}

\item L1 (least absolute deviations regression): $$L(w, \hat{w}) = ||w - \hat{w}||_1$$

\item L2 (least squares regression): $$L(w, \hat{w}) = ||w - \hat{w}||_2^2$$

\item SmoothL1: L1 loss is not strongly convex thus the solution is less stable  compared to L2 loss, while L2 loss is sensitive for the outliers and vulnerable to exploding gradients\cite{koenker2001quantile,girshick2015fast}. SmoothL1 loss, also known as the Huber loss,  is a combination of L1 and L2 loss which enjoys the advantages from both of them \cite{girshick2015fast}. It is  implemented in PyTorch \cite{paszke2017automatic}:

    \[
   L(w, \hat{w}) =
   \begin{cases}
      0.5  (w - \hat{w})^2,& \for |||w - \hat{w}||_1 < 1 \\
      ||w - \hat{w}||_1, & \for |||w - \hat{w}||_1 >= 1
   \end{cases}
   \]

\end{itemize}

\subsubsection{Unweighted Graph}

We also considered the unweighted graph. The only difference from the weighted graph in \ref{sec:weight} is that, instead of using the weight of the link, we dichotomized the weighted edge into a unweighted binary link. For instance, we can define there is a link between user $u_i$ and domain $d_j$, if the user reshared some link from domain $d_j$ more than twice in the last week. In this manner, we set the target $w_{u_i, d_j}$ to be a binary variable, and the output from the neural network $\hat{w}_{u_i, d_j}$ is bounded in $[0, 1]$, which can be interpreted as the probability of the existence of a link within this user-domain pair. As the target in this setting is binary, we considered the following loss functions:

\begin{itemize}

\item CrossEntropy: this is one of the most common loss functions used in classification:

  $$L(w_{u_i, d_j}, \hat{w}_{u_i, d_j}) = (1 - w_{u_i, d_j})\log(1-\vec{w}_{u_i, d_j})
  + w_{u_i, d_j} \log( \hat{w}_{u_i, d_j}) $$

\end{itemize}

We also consider the embedding-distance based loss functions. These functions does not inputing the embedding into a new block of neural network. Instead, it  only relies on the distance between the user and the domain embedding $e_u, e_d$, and \textbf{binary indicator} of the existence of the edge $w_{u, d}$.

\begin{itemize}

\item Contrastive Loss: this loss decreases the energy  of like pairs and increase the energy  of unlike pairs  \cite{koch2015siamese,chopra2005learning}. Here we define the energy as one minus the output of the graph regularization building block. Recall that the output of the graph regularization building block represents the predicted existence of the edge between the given user-domain pair. We simply set the margin $m$ to be $0.2$.

  \begin{eqnarray*}
    \text{dist} &=&||e_{u_i} - e_{d_j}||_2\\
   L(w_{u_i, d_j}, e_{u_i}, e_{d_j}) &=& w \cdot \text{dist}^2  +  (1 - w)  [\max(0,m - \text{dist})]^2
\end{eqnarray*}

\item CosineEmbed: we consider the cosine embedding loss implemented in PyTorch \cite{paszke2017automatic}:
    \[
   L(w_{u_i, d_j}, e_{u_i}, e_{d_j}) =
   \begin{cases}
     1 - \cos(e_{u_i}, e_{d_j}),& \for | w_{u_i, d_j} = 1 \\
      \cos(e_{u_i}, e_{d_j}) & \for | w_{u_i, d_j} = 0
   \end{cases}
   \]

 \item L1Embed: we also consider the L1 and L2  distance metric used in neural graphical machines \cite{bui2017neural}:

$$ L(w_{u_i, d_j},  e_{u_i}, e_{d_j})  = w_i || e_{u_i} - e_{d_j}||_1$$

\item L2Embed:

$$ L(w_{u_i, d_j},  e_{u_i}, e_{d_j}) = w_i || e_{u_i} - e_{d_j}||_2^2 $$

\end{itemize}

\begin{table}[ht]
  \caption{The predictive performance for HELP with different unsupervised loss on testing set for dimension\#2 domain label. All the values are in $10^{-2}$ scale.}
\begin{center}
  \begin{tabular}{c|c|c
    }
    Loss & AUROC & AUPRC 
    \\ \hline \hline
    \textbf{Contrastive} & 96.3 & 82.9 
    \\ \hline \hline
    CosineEmbed & 95.6 & 82.1
    \\ 
    L1Embed & 95.2 & 81.6 
    \\
    L2Embed & 95.3 & 81.4 
    \\
    \hline \hline
    L1 & 96.0 & 82.7 
    \\
    L2 & 95.8 & 82.1 
    \\
    SmoothL1 & 95.9 & 82.5 
    \\
    \hline  \hline
    CrossEntropy & 96.1  & 82.8 
    \\
    \hline \hline
    MLP & 95.6 & 81.6  
    \\
  \end{tabular}
  \label{tab:loss}
  \caption{The performance for different loss function when considering dimension\#2 label.}
\end{center}
\end{table}

Table \ref{tab:loss} shows the performance of the HELP model with different unsupervised loss. Among all the loss choices, the HELP with contrastive loss achieves both the best performance for AUROC and AUPRC. The other three embedding based loss, CosineEmbed, L1Embed and L2Embed, achieves worse performance. This may be explained by the flexible distance evaluation. For contrastive loss we used here, we generate the distance from a feed forward neural network with the embedding from both user and domain as input, instead of a fixed commonly used distance metric like cosine distance. This makes the distance selection more flexible.

In addition, we observe the L1Embed and L2Embed is noticeably worse than CosineEmbed and Contrastive, and they does not show any improvement compared to simple MLP. This might due to the L1/L2 losses only ``pull'' the connected pair closer, while both CosineEmbed and Contrastive loss not only ``pull'' the connected pair closer, but also ``push'' the unconnected pair farther away, and therefore improves the learning of the embedding.

For the classification based loss (L1, L2, SmoothL1, and CrossEntropy), we observed all of them has improvement compared to the benchmark MLP. The L2 loss has slightly worse performance compared th L1 and SmoothL1, a combination of L1 and L2 loss. This might due to some extreme weight in the edge, which make too strong impact when training the network. Furthermore, when  edges are treated unweighted  by thresholding weighted edge, the performance is slightly improved. Similar to previous explanation, we believe such discretization improve the performance by avoid the outliers in the edge weights. A potential solution of it would be truncate the loss for unweighted edge, and we leave it for future work.

\section{Unsupervised Learning}

As discussed above, we do not have explicit label for each user. However, we define some ad-hoc labels for each user to assess the effectiveness of the user embedding, a side-produce in the HELP model. 

\subsection{Visualization of Embedding}

We visualize the embedding for users, which is the side-product of the HELP model.

\textbf{To avoid information leakage/over-fitting during the training, we generate the graph with the interactions 1 week after the training data for visualization. In other word the graph is generated by the interactions between user and domain from 10/27/2017 to 11/03/2017. In addition, the user features/domain labels in our visualization are also collected one week after the collecting date of the experiment data.}

We investigate and visualize the users that might be ``vulnerable'' to dimension\#2 domains, which we defined as the active users with frequent interaction with some dimension\#2 domains. To be more specific:

\begin{itemize}
\item For each type of interaction (e.g. clicking the link), we first select the  users that have more than 5 such interactions during the whole evaluating week as active users.
\item Among such users, if the user is more than 5 such interaction with domains that labeled as dimension\#2 domain  in one week, we define this user as a vulnerable (positive) user.
\item In visualization, we use the red (positive) nodes to represent the  vulnerable users, while using blue (negative) nodes for the remaining active users.
\item As there are much less positive samples, we down sampled the negative samples to relative same size as positive samples.
\end{itemize}

In this section, we studied the five different interaction types, including:

\begin{itemize}
\item Click: clicking of the link.
\item Reshare: resharing the link.
\item Wow: Clicking the Wow button for the link.
\item Angry: Clicking the Angry button for the link.
  \end{itemize}

We compared the user embedding generated from the HELP, and the raw features. We use t-SNE to reduce the dimension to 2, while maintaining the Euclidean distance between nodes for both raw features \cite{maaten2008visualizing}  and the generated embedding from the HELP. We simply used the t-SNE function with default parameter in sklearn  \cite{scikit-learn}.
Then we plot each node on 2-D space, with color represents if the node is a vulnerable use or not.

\begin{figure}[H]
    \centering
    \begin{subfigure}[b]{0.2\textwidth}
      \includegraphics[width=\textwidth]{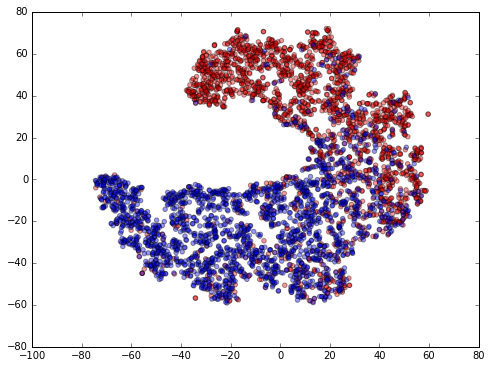}
    \end{subfigure}
     \begin{subfigure}[b]{0.2\textwidth}
        \includegraphics[width=\textwidth]{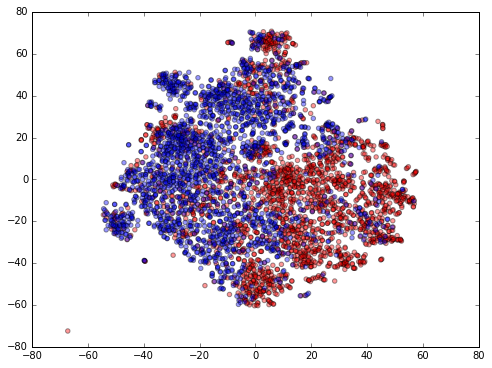}
    \end{subfigure}
     \caption{Click. The left figure is for the embedding from the HELP; the right figure is for the raw features.}
     \label{fig:click}
\end{figure}

\begin{figure}[H]
    \centering
    \begin{subfigure}[b]{0.2\textwidth}
      \includegraphics[width=\textwidth]{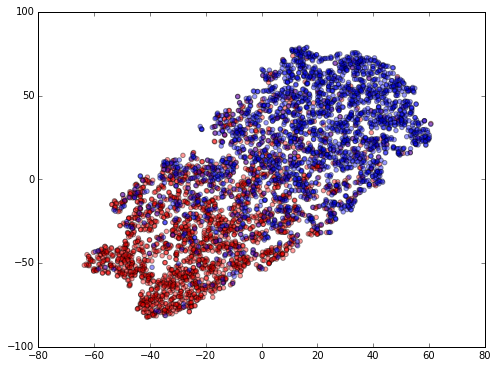}
    \end{subfigure}
     \begin{subfigure}[b]{0.2\textwidth}
        \includegraphics[width=\textwidth]{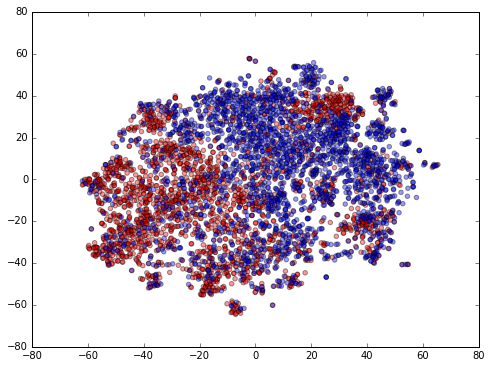}
    \end{subfigure}
     \caption{Reshare. The left figure is for the embedding from the HELP; the right figure is for the raw features.}
     \label{fig:reshare}
\end{figure}
         
Figure \ref{fig:click} and \ref{fig:reshare} shows the visualization comparison for Click and Reshare. For both Click and Reshare, we can observe a passable pattern for the separation of blue/red nodes even for the raw features. Though most of the blue nodes are on the one side, there are still many regions that blue and red nodes are mixed. However, the embedding from the HELP further pulled the users of different type further away. We can observe very clear separation boundary for two type of users.

\begin{figure}[H]
    \centering
    \begin{subfigure}[b]{0.2\textwidth}
      \includegraphics[width=\textwidth]{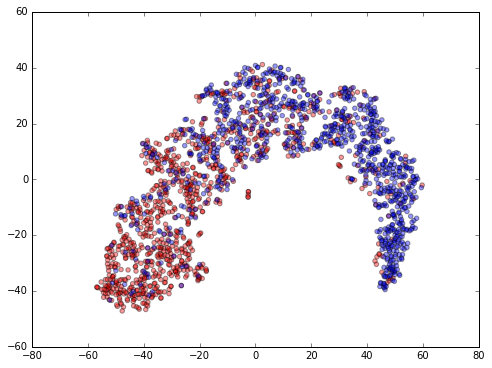}
    \end{subfigure}
     \begin{subfigure}[b]{0.2\textwidth}
        \includegraphics[width=\textwidth]{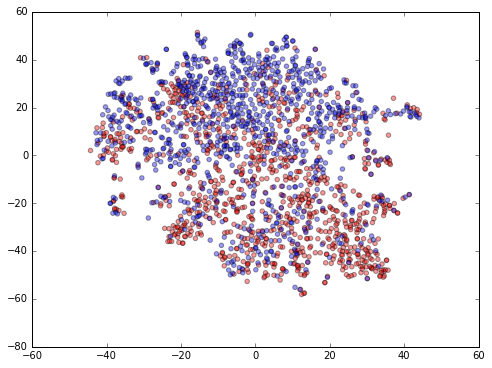}
    \end{subfigure}
     \caption{Wow. The left figure is for the embedding from the HELP; the right figure is for the raw features.}
     \label{fig:wow}
\end{figure}

\begin{figure}[H]
    \centering
    \begin{subfigure}[b]{0.2\textwidth}
      \includegraphics[width=\textwidth]{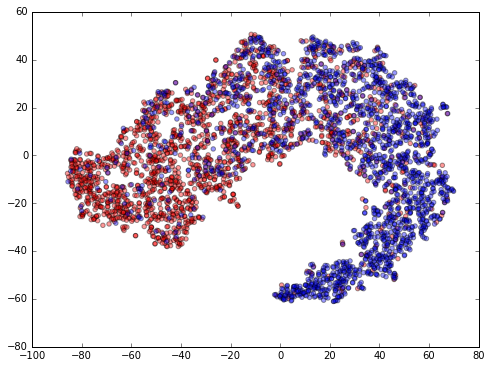}
    \end{subfigure}
     \begin{subfigure}[b]{0.2\textwidth}
        \includegraphics[width=\textwidth]{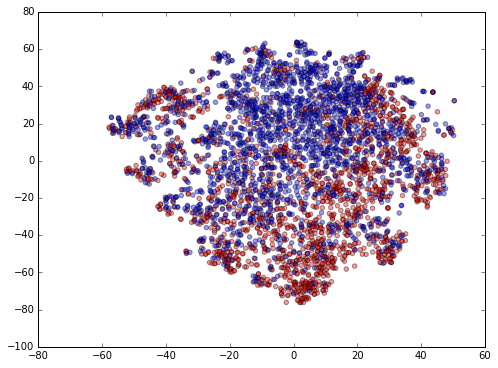}
    \end{subfigure}
     \caption{Angry. The left figure is for the embedding from the HELP; the right figure is for the raw features.}
     \label{fig:angry}
\end{figure}

Figure \ref{fig:wow} and \ref{fig:angry} shows the visualization comparison for Wow and Angry. For these interaction types, the raw features did a bad job in separating two different type of users. However, the embedding from the HELP still achieves satisfactory performance in separating two type of users. 

In conclusion, the HELP generates embedding for users as a side-product. Our visualization results suggest such user-level embedding can help other tasks, like user-level clustering.

\section{Discussion}

In this work, we propose HELP, a graph-based semi-supervised deep learning method for graphs with heterogeneous type of nodes. We demonstrate its performance with several domain classification tasks for News Feed in Facebook. One potential future direction is  multi-tasks prediction to predict different type of label simultaneously: we can extend the network architecture by stacking a multiple-output prediction layer on the second last layer, which output a vector of probability for multiple label dimensions. This can be done by extending the supervised loss with the multi-label loss. It has following benefits: first the model size can be compressed as we only need to train one model for multi-labels. Second, the embedding generated in this network contains information for different label type, thus is more informative and can be potentially used as a general ``integrity reputation embedding'' for a domain.

Another interesting direction is allowing different types of edges between nodes. In our experiments, we only consider the ``resharing interaction'' edges. Different type of edge can be included to further improve the performance of the semi-supervised approach. In addition, we may use weighted combination of multiple interaction types as the weight in graph.

We directly concatenated two embedding and then feed it into a building block to estimate the similarity for each pair. Instead of concatenating, several different approaches can be applied to combine the embedding of the domain-user pair, which may further improve the performance of the HELP. For example we may consider the element-wise product/difference of two embedding vectors.

There are also several minor changes may further improve the performance of the HELP. We set margin $m=0.2$ in an  ad-hoc manner for the contrastive loss, which can be further investigated. We can also extend the EmbedL1/EmbedL2 loss by imitating the contrastive loss that including penalization for the unconnected pair with small distance. Due to the limited space, we leave this as our future work.

\section{Acknowledgements}

Authors sincerely appreciate the Facebook News Feed team for the help during the project and the insightful feedback.

\newpage
\bibliographystyle{ACM-Reference-Format}
\bibliography{help} 


\begin{thebibliography}{00}


\ifx \showCODEN    \undefined \def \showCODEN     #1{\unskip}     \fi
\ifx \showDOI      \undefined \def \showDOI       #1{{\tt DOI:}\penalty0{#1}\ }
  \fi
\ifx \showISBNx    \undefined \def \showISBNx     #1{\unskip}     \fi
\ifx \showISBNxiii \undefined \def \showISBNxiii  #1{\unskip}     \fi
\ifx \showISSN     \undefined \def \showISSN      #1{\unskip}     \fi
\ifx \showLCCN     \undefined \def \showLCCN      #1{\unskip}     \fi
\ifx \shownote     \undefined \def \shownote      #1{#1}          \fi
\ifx \showarticletitle \undefined \def \showarticletitle #1{#1}   \fi
\ifx \showURL      \undefined \def \showURL       #1{#1}          \fi
\providecommand\bibfield[2]{#2}
\providecommand\bibinfo[2]{#2}
\providecommand\natexlab[1]{#1}
\providecommand\showeprint[2][]{arXiv:#2}

\bibitem[\protect\citeauthoryear{Al~Hasan, Chaoji, Salem, and Zaki}{Al~Hasan
  et~al\mbox{.}}{2006}]%
        {al2006link}
\bibfield{author}{\bibinfo{person}{Mohammad Al~Hasan}, \bibinfo{person}{Vineet
  Chaoji}, \bibinfo{person}{Saeed Salem}, {and} \bibinfo{person}{Mohammed
  Zaki}.} \bibinfo{year}{2006}\natexlab{}.
\newblock \showarticletitle{Link prediction using supervised learning}. In
  \bibinfo{booktitle}{{\em SDM06: workshop on link analysis, counter-terrorism
  and security}}.
\newblock


\bibitem[\protect\citeauthoryear{Babu, A., and Zhang}{Babu
  et~al\mbox{.}}{2017}]%
        {feedfyicb}
\bibfield{author}{\bibinfo{person}{A. Babu}, \bibinfo{person}{Liu A.}, {and}
  \bibinfo{person}{J Zhang}.} \bibinfo{year}{2017}\natexlab{}.
\newblock \bibinfo{title}{News Feed FYI: New Updates to Reduce Clickbait
  Headlines}.
\newblock
  \bibinfo{howpublished}{\url{https://newsroom.fb.com/news/2017/05/news-feed-fyi-new-updates-to-reduce-clickbait-headlines/}}.
    (\bibinfo{year}{2017}).
\newblock


\bibitem[\protect\citeauthoryear{Backstrom}{Backstrom}{2016}]%
        {newsfeedlars}
\bibfield{author}{\bibinfo{person}{Lars Backstrom}.}
  \bibinfo{year}{2016}\natexlab{}.
\newblock \showarticletitle{Serving a Billion Personalized News Feeds}. In
  \bibinfo{booktitle}{{\em Proceedings of the Ninth {ACM} International
  Conference on Web Search and Data Mining, San Francisco, CA, USA, February
  22-25, 2016}}. \bibinfo{pages}{469}.
\newblock
\showDOI{%
\url{http://dx.doi.org/10.1145/2835776.2835848}}


\bibitem[\protect\citeauthoryear{Belkin, Niyogi, and Sindhwani}{Belkin
  et~al\mbox{.}}{2006}]%
        {belkin2006manifold}
\bibfield{author}{\bibinfo{person}{Mikhail Belkin}, \bibinfo{person}{Partha
  Niyogi}, {and} \bibinfo{person}{Vikas Sindhwani}.}
  \bibinfo{year}{2006}\natexlab{}.
\newblock \showarticletitle{Manifold regularization: A geometric framework for
  learning from labeled and unlabeled examples}.
\newblock \bibinfo{journal}{{\em Journal of machine learning research\/}}
  \bibinfo{volume}{7}, \bibinfo{number}{Nov} (\bibinfo{year}{2006}),
  \bibinfo{pages}{2399--2434}.
\newblock


\bibitem[\protect\citeauthoryear{Bui, Ravi, and Ramavajjala}{Bui
  et~al\mbox{.}}{2017}]%
        {bui2017neural}
\bibfield{author}{\bibinfo{person}{Thang~D Bui}, \bibinfo{person}{Sujith Ravi},
  {and} \bibinfo{person}{Vivek Ramavajjala}.} \bibinfo{year}{2017}\natexlab{}.
\newblock \showarticletitle{Neural Graph Machines: Learning Neural Networks
  Using Graphs}.
\newblock \bibinfo{journal}{{\em arXiv preprint arXiv:1703.04818\/}}
  (\bibinfo{year}{2017}).
\newblock


\bibitem[\protect\citeauthoryear{Chang, Han, Tang, Qi, Aggarwal, and
  Huang}{Chang et~al\mbox{.}}{2015}]%
        {chang2015heterogeneous}
\bibfield{author}{\bibinfo{person}{Shiyu Chang}, \bibinfo{person}{Wei Han},
  \bibinfo{person}{Jiliang Tang}, \bibinfo{person}{Guo-Jun Qi},
  \bibinfo{person}{Charu~C Aggarwal}, {and} \bibinfo{person}{Thomas~S Huang}.}
  \bibinfo{year}{2015}\natexlab{}.
\newblock \showarticletitle{Heterogeneous network embedding via deep
  architectures}. In \bibinfo{booktitle}{{\em Proceedings of the 21th ACM
  SIGKDD International Conference on Knowledge Discovery and Data Mining}}.
  ACM, \bibinfo{pages}{119--128}.
\newblock


\bibitem[\protect\citeauthoryear{Chopra, Hadsell, and LeCun}{Chopra
  et~al\mbox{.}}{2005}]%
        {chopra2005learning}
\bibfield{author}{\bibinfo{person}{Sumit Chopra}, \bibinfo{person}{Raia
  Hadsell}, {and} \bibinfo{person}{Yann LeCun}.}
  \bibinfo{year}{2005}\natexlab{}.
\newblock \showarticletitle{Learning a similarity metric discriminatively, with
  application to face verification}. In \bibinfo{booktitle}{{\em Computer
  Vision and Pattern Recognition, 2005. CVPR 2005. IEEE Computer Society
  Conference on}}, Vol.~\bibinfo{volume}{1}. IEEE, \bibinfo{pages}{539--546}.
\newblock


\bibitem[\protect\citeauthoryear{Dong, Chawla, and Swami}{Dong
  et~al\mbox{.}}{2017}]%
        {dong2017metapath2vec}
\bibfield{author}{\bibinfo{person}{Yuxiao Dong}, \bibinfo{person}{Nitesh~V
  Chawla}, {and} \bibinfo{person}{Ananthram Swami}.}
  \bibinfo{year}{2017}\natexlab{}.
\newblock \showarticletitle{metapath2vec: Scalable representation learning for
  heterogeneous networks}. In \bibinfo{booktitle}{{\em Proceedings of the 23rd
  ACM SIGKDD International Conference on Knowledge Discovery and Data Mining}}.
  ACM, \bibinfo{pages}{135--144}.
\newblock


\bibitem[\protect\citeauthoryear{Girshick}{Girshick}{2015}]%
        {girshick2015fast}
\bibfield{author}{\bibinfo{person}{Ross Girshick}.}
  \bibinfo{year}{2015}\natexlab{}.
\newblock \showarticletitle{Fast r-cnn}. In \bibinfo{booktitle}{{\em
  Proceedings of the IEEE international conference on computer vision}}.
  \bibinfo{pages}{1440--1448}.
\newblock


\bibitem[\protect\citeauthoryear{Goldberg and Levy}{Goldberg and Levy}{2014}]%
        {goldberg2014word2vec}
\bibfield{author}{\bibinfo{person}{Yoav Goldberg} {and} \bibinfo{person}{Omer
  Levy}.} \bibinfo{year}{2014}\natexlab{}.
\newblock \showarticletitle{word2vec Explained: deriving Mikolov et al.'s
  negative-sampling word-embedding method}.
\newblock \bibinfo{journal}{{\em arXiv preprint arXiv:1402.3722\/}}
  (\bibinfo{year}{2014}).
\newblock


\bibitem[\protect\citeauthoryear{Grover and Leskovec}{Grover and
  Leskovec}{2016}]%
        {grover2016node2vec}
\bibfield{author}{\bibinfo{person}{Aditya Grover} {and} \bibinfo{person}{Jure
  Leskovec}.} \bibinfo{year}{2016}\natexlab{}.
\newblock \showarticletitle{node2vec: Scalable feature learning for networks}.
  In \bibinfo{booktitle}{{\em Proceedings of the 22nd ACM SIGKDD international
  conference on Knowledge discovery and data mining}}. ACM,
  \bibinfo{pages}{855--864}.
\newblock


\bibitem[\protect\citeauthoryear{Jacob, Denoyer, and Gallinari}{Jacob
  et~al\mbox{.}}{2014}]%
        {jacob2014learning}
\bibfield{author}{\bibinfo{person}{Yann Jacob}, \bibinfo{person}{Ludovic
  Denoyer}, {and} \bibinfo{person}{Patrick Gallinari}.}
  \bibinfo{year}{2014}\natexlab{}.
\newblock \showarticletitle{Learning latent representations of nodes for
  classifying in heterogeneous social networks}. In \bibinfo{booktitle}{{\em
  Proceedings of the 7th ACM international conference on Web search and data
  mining}}. ACM, \bibinfo{pages}{373--382}.
\newblock


\bibitem[\protect\citeauthoryear{Ju, Bibaut, and van~der Laan}{Ju
  et~al\mbox{.}}{2017}]%
        {ju2017relative}
\bibfield{author}{\bibinfo{person}{Cheng Ju}, \bibinfo{person}{Aur{\'e}lien
  Bibaut}, {and} \bibinfo{person}{Mark~J van~der Laan}.}
  \bibinfo{year}{2017}\natexlab{}.
\newblock \showarticletitle{The Relative Performance of Ensemble Methods with
  Deep Convolutional Neural Networks for Image Classification}.
\newblock \bibinfo{journal}{{\em arXiv preprint arXiv:1704.01664\/}}
  (\bibinfo{year}{2017}).
\newblock


\bibitem[\protect\citeauthoryear{Kingma and Ba}{Kingma and Ba}{2014}]%
        {kingma2014adam}
\bibfield{author}{\bibinfo{person}{Diederik Kingma} {and}
  \bibinfo{person}{Jimmy Ba}.} \bibinfo{year}{2014}\natexlab{}.
\newblock \showarticletitle{Adam: A method for stochastic optimization}.
\newblock \bibinfo{journal}{{\em arXiv preprint arXiv:1412.6980\/}}
  (\bibinfo{year}{2014}).
\newblock


\bibitem[\protect\citeauthoryear{Koch, Zemel, and Salakhutdinov}{Koch
  et~al\mbox{.}}{2015}]%
        {koch2015siamese}
\bibfield{author}{\bibinfo{person}{Gregory Koch}, \bibinfo{person}{Richard
  Zemel}, {and} \bibinfo{person}{Ruslan Salakhutdinov}.}
  \bibinfo{year}{2015}\natexlab{}.
\newblock \showarticletitle{Siamese neural networks for one-shot image
  recognition}. In \bibinfo{booktitle}{{\em ICML Deep Learning Workshop}},
  Vol.~\bibinfo{volume}{2}.
\newblock


\bibitem[\protect\citeauthoryear{Koenker and Hallock}{Koenker and
  Hallock}{2001}]%
        {koenker2001quantile}
\bibfield{author}{\bibinfo{person}{Roger Koenker} {and}
  \bibinfo{person}{Kevin~F Hallock}.} \bibinfo{year}{2001}\natexlab{}.
\newblock \showarticletitle{Quantile regression}.
\newblock \bibinfo{journal}{{\em Journal of economic perspectives\/}}
  \bibinfo{volume}{15}, \bibinfo{number}{4} (\bibinfo{year}{2001}),
  \bibinfo{pages}{143--156}.
\newblock


\bibitem[\protect\citeauthoryear{Lada, Li, and Ding}{Lada
  et~al\mbox{.}}{2017}]%
        {feedfyihq}
\bibfield{author}{\bibinfo{person}{A. Lada}, \bibinfo{person}{J. Li}, {and}
  \bibinfo{person}{S. Ding}.} \bibinfo{year}{2017}\natexlab{}.
\newblock \bibinfo{title}{News Feed FYI: New Signals to Show You More Authentic
  and Timely Stories}.
\newblock
  \bibinfo{howpublished}{\url{https://newsroom.fb.com/news/2017/01/news-feed-fyi-new-signals-to-show-you-more-authentic-and-timely-stories/}}.
    (\bibinfo{year}{2017}).
\newblock


\bibitem[\protect\citeauthoryear{Liben-Nowell and Kleinberg}{Liben-Nowell and
  Kleinberg}{2007}]%
        {liben2007link}
\bibfield{author}{\bibinfo{person}{David Liben-Nowell} {and}
  \bibinfo{person}{Jon Kleinberg}.} \bibinfo{year}{2007}\natexlab{}.
\newblock \showarticletitle{The link-prediction problem for social networks}.
\newblock \bibinfo{journal}{{\em Journal of the American society for
  information science and technology\/}} \bibinfo{volume}{58},
  \bibinfo{number}{7} (\bibinfo{year}{2007}), \bibinfo{pages}{1019--1031}.
\newblock


\bibitem[\protect\citeauthoryear{Lin and S.}{Lin and S.}{2017}]%
        {feedfyiaf}
\bibfield{author}{\bibinfo{person}{J. Lin} {and} \bibinfo{person}{Guo S.}}
  \bibinfo{year}{2017}\natexlab{}.
\newblock \bibinfo{title}{News Feed FYI: Reducing Links to Low-Quality Web Page
  Experiences}.
\newblock
  \bibinfo{howpublished}{\url{https://newsroom.fb.com/news/2017/05/reducing-links-to-low-quality-web-page-experiences/}}.
    (\bibinfo{year}{2017}).
\newblock


\bibitem[\protect\citeauthoryear{L{\"u} and Zhou}{L{\"u} and Zhou}{2011}]%
        {lu2011link}
\bibfield{author}{\bibinfo{person}{Linyuan L{\"u}} {and} \bibinfo{person}{Tao
  Zhou}.} \bibinfo{year}{2011}\natexlab{}.
\newblock \showarticletitle{Link prediction in complex networks: A survey}.
\newblock \bibinfo{journal}{{\em Physica A: Statistical Mechanics and its
  Applications\/}} \bibinfo{volume}{390}, \bibinfo{number}{6}
  (\bibinfo{year}{2011}), \bibinfo{pages}{1150--1170}.
\newblock


\bibitem[\protect\citeauthoryear{Maaten and Hinton}{Maaten and Hinton}{2008}]%
        {maaten2008visualizing}
\bibfield{author}{\bibinfo{person}{Laurens van~der Maaten} {and}
  \bibinfo{person}{Geoffrey Hinton}.} \bibinfo{year}{2008}\natexlab{}.
\newblock \showarticletitle{Visualizing data using t-SNE}.
\newblock \bibinfo{journal}{{\em Journal of Machine Learning Research\/}}
  \bibinfo{volume}{9}, \bibinfo{number}{Nov} (\bibinfo{year}{2008}),
  \bibinfo{pages}{2579--2605}.
\newblock


\bibitem[\protect\citeauthoryear{Mikolov, Sutskever, Chen, Corrado, and
  Dean}{Mikolov et~al\mbox{.}}{2013}]%
        {mikolov2013distributed}
\bibfield{author}{\bibinfo{person}{Tomas Mikolov}, \bibinfo{person}{Ilya
  Sutskever}, \bibinfo{person}{Kai Chen}, \bibinfo{person}{Greg~S Corrado},
  {and} \bibinfo{person}{Jeff Dean}.} \bibinfo{year}{2013}\natexlab{}.
\newblock \showarticletitle{Distributed representations of words and phrases
  and their compositionality}. In \bibinfo{booktitle}{{\em Advances in neural
  information processing systems}}. \bibinfo{pages}{3111--3119}.
\newblock


\bibitem[\protect\citeauthoryear{Newman}{Newman}{2006}]%
        {newman2006finding}
\bibfield{author}{\bibinfo{person}{Mark~EJ Newman}.}
  \bibinfo{year}{2006}\natexlab{}.
\newblock \showarticletitle{Finding community structure in networks using the
  eigenvectors of matrices}.
\newblock \bibinfo{journal}{{\em Physical review E\/}} \bibinfo{volume}{74},
  \bibinfo{number}{3} (\bibinfo{year}{2006}), \bibinfo{pages}{036104}.
\newblock


\bibitem[\protect\citeauthoryear{Ng, Jordan, Weiss, et~al\mbox{.}}{Ng
  et~al\mbox{.}}{2002}]%
        {ng2002spectral}
\bibfield{author}{\bibinfo{person}{Andrew~Y Ng}, \bibinfo{person}{Michael~I
  Jordan}, \bibinfo{person}{Yair Weiss}, {and} \bibinfo{person}{others}.}
  \bibinfo{year}{2002}\natexlab{}.
\newblock \showarticletitle{On spectral clustering: Analysis and an algorithm}.
\newblock \bibinfo{journal}{{\em Advances in neural information processing
  systems\/}}  \bibinfo{volume}{2} (\bibinfo{year}{2002}),
  \bibinfo{pages}{849--856}.
\newblock


\bibitem[\protect\citeauthoryear{Paszke, Gross, Chintala, Chanan, Yang, DeVito,
  Lin, Desmaison, Antiga, and Lerer}{Paszke et~al\mbox{.}}{2017}]%
        {paszke2017automatic}
\bibfield{author}{\bibinfo{person}{Adam Paszke}, \bibinfo{person}{Sam Gross},
  \bibinfo{person}{Soumith Chintala}, \bibinfo{person}{Gregory Chanan},
  \bibinfo{person}{Edward Yang}, \bibinfo{person}{Zachary DeVito},
  \bibinfo{person}{Zeming Lin}, \bibinfo{person}{Alban Desmaison},
  \bibinfo{person}{Luca Antiga}, {and} \bibinfo{person}{Adam Lerer}.}
  \bibinfo{year}{2017}\natexlab{}.
\newblock \showarticletitle{Automatic differentiation in PyTorch}.
\newblock  (\bibinfo{year}{2017}).
\newblock


\bibitem[\protect\citeauthoryear{Pedregosa, Varoquaux, Gramfort, Michel,
  Thirion, Grisel, Blondel, Prettenhofer, Weiss, Dubourg, Vanderplas, Passos,
  Cournapeau, Brucher, Perrot, and Duchesnay}{Pedregosa et~al\mbox{.}}{2011}]%
        {scikit-learn}
\bibfield{author}{\bibinfo{person}{F. Pedregosa}, \bibinfo{person}{G.
  Varoquaux}, \bibinfo{person}{A. Gramfort}, \bibinfo{person}{V. Michel},
  \bibinfo{person}{B. Thirion}, \bibinfo{person}{O. Grisel},
  \bibinfo{person}{M. Blondel}, \bibinfo{person}{P. Prettenhofer},
  \bibinfo{person}{R. Weiss}, \bibinfo{person}{V. Dubourg}, \bibinfo{person}{J.
  Vanderplas}, \bibinfo{person}{A. Passos}, \bibinfo{person}{D. Cournapeau},
  \bibinfo{person}{M. Brucher}, \bibinfo{person}{M. Perrot}, {and}
  \bibinfo{person}{E. Duchesnay}.} \bibinfo{year}{2011}\natexlab{}.
\newblock \showarticletitle{Scikit-learn: Machine Learning in {P}ython}.
\newblock \bibinfo{journal}{{\em Journal of Machine Learning Research\/}}
  \bibinfo{volume}{12} (\bibinfo{year}{2011}), \bibinfo{pages}{2825--2830}.
\newblock


\bibitem[\protect\citeauthoryear{Perozzi, Al-Rfou, and Skiena}{Perozzi
  et~al\mbox{.}}{2014}]%
        {perozzi2014deepwalk}
\bibfield{author}{\bibinfo{person}{Bryan Perozzi}, \bibinfo{person}{Rami
  Al-Rfou}, {and} \bibinfo{person}{Steven Skiena}.}
  \bibinfo{year}{2014}\natexlab{}.
\newblock \showarticletitle{Deepwalk: Online learning of social
  representations}. In \bibinfo{booktitle}{{\em Proceedings of the 20th ACM
  SIGKDD international conference on Knowledge discovery and data mining}}.
  ACM, \bibinfo{pages}{701--710}.
\newblock


\bibitem[\protect\citeauthoryear{Saito and Rehmsmeier}{Saito and
  Rehmsmeier}{2015}]%
        {saito2015precision}
\bibfield{author}{\bibinfo{person}{Takaya Saito} {and} \bibinfo{person}{Marc
  Rehmsmeier}.} \bibinfo{year}{2015}\natexlab{}.
\newblock \showarticletitle{The precision-recall plot is more informative than
  the ROC plot when evaluating binary classifiers on imbalanced datasets}.
\newblock \bibinfo{journal}{{\em PloS one\/}} \bibinfo{volume}{10},
  \bibinfo{number}{3} (\bibinfo{year}{2015}), \bibinfo{pages}{e0118432}.
\newblock


\bibitem[\protect\citeauthoryear{Tang, Qu, Wang, Zhang, Yan, and Mei}{Tang
  et~al\mbox{.}}{2015}]%
        {tang2015line}
\bibfield{author}{\bibinfo{person}{Jian Tang}, \bibinfo{person}{Meng Qu},
  \bibinfo{person}{Mingzhe Wang}, \bibinfo{person}{Ming Zhang},
  \bibinfo{person}{Jun Yan}, {and} \bibinfo{person}{Qiaozhu Mei}.}
  \bibinfo{year}{2015}\natexlab{}.
\newblock \showarticletitle{Line: Large-scale information network embedding}.
  In \bibinfo{booktitle}{{\em Proceedings of the 24th International Conference
  on World Wide Web}}. International World Wide Web Conferences Steering
  Committee, \bibinfo{pages}{1067--1077}.
\newblock


\bibitem[\protect\citeauthoryear{Tang and Liu}{Tang and Liu}{2009a}]%
        {tang2009relational}
\bibfield{author}{\bibinfo{person}{Lei Tang} {and} \bibinfo{person}{Huan Liu}.}
  \bibinfo{year}{2009}\natexlab{a}.
\newblock \showarticletitle{Relational learning via latent social dimensions}.
  In \bibinfo{booktitle}{{\em Proceedings of the 15th ACM SIGKDD international
  conference on Knowledge discovery and data mining}}. ACM,
  \bibinfo{pages}{817--826}.
\newblock


\bibitem[\protect\citeauthoryear{Tang and Liu}{Tang and Liu}{2009b}]%
        {tang2009scalable}
\bibfield{author}{\bibinfo{person}{Lei Tang} {and} \bibinfo{person}{Huan Liu}.}
  \bibinfo{year}{2009}\natexlab{b}.
\newblock \showarticletitle{Scalable learning of collective behavior based on
  sparse social dimensions}. In \bibinfo{booktitle}{{\em Proceedings of the
  18th ACM conference on Information and knowledge management}}. ACM,
  \bibinfo{pages}{1107--1116}.
\newblock


\bibitem[\protect\citeauthoryear{Tang and Liu}{Tang and Liu}{2011}]%
        {tang2011leveraging}
\bibfield{author}{\bibinfo{person}{Lei Tang} {and} \bibinfo{person}{Huan Liu}.}
  \bibinfo{year}{2011}\natexlab{}.
\newblock \showarticletitle{Leveraging social media networks for
  classification}.
\newblock \bibinfo{journal}{{\em Data Mining and Knowledge Discovery\/}}
  \bibinfo{volume}{23}, \bibinfo{number}{3} (\bibinfo{year}{2011}),
  \bibinfo{pages}{447--478}.
\newblock


\bibitem[\protect\citeauthoryear{Yang, Cohen, and Salakhudinov}{Yang
  et~al\mbox{.}}{2016}]%
        {yang2016revisiting}
\bibfield{author}{\bibinfo{person}{Zhilin Yang}, \bibinfo{person}{William
  Cohen}, {and} \bibinfo{person}{Ruslan Salakhudinov}.}
  \bibinfo{year}{2016}\natexlab{}.
\newblock \showarticletitle{Revisiting Semi-Supervised Learning with Graph
  Embeddings}. In \bibinfo{booktitle}{{\em International Conference on Machine
  Learning}}. \bibinfo{pages}{40--48}.
\newblock


\bibitem[\protect\citeauthoryear{Zhou, Bousquet, Lal, Weston, and
  Sch{\"o}lkopf}{Zhou et~al\mbox{.}}{2004}]%
        {zhou2004learning}
\bibfield{author}{\bibinfo{person}{Denny Zhou}, \bibinfo{person}{Olivier
  Bousquet}, \bibinfo{person}{Thomas~N Lal}, \bibinfo{person}{Jason Weston},
  {and} \bibinfo{person}{Bernhard Sch{\"o}lkopf}.}
  \bibinfo{year}{2004}\natexlab{}.
\newblock \showarticletitle{Learning with local and global consistency}. In
  \bibinfo{booktitle}{{\em Advances in neural information processing systems}}.
  \bibinfo{pages}{321--328}.
\newblock


\bibitem[\protect\citeauthoryear{Zhu and Ghahramani}{Zhu and
  Ghahramani}{2002}]%
        {zhu2002learning}
\bibfield{author}{\bibinfo{person}{Xiaojin Zhu} {and} \bibinfo{person}{Zoubin
  Ghahramani}.} \bibinfo{year}{2002}\natexlab{}.
\newblock \showarticletitle{Learning from labeled and unlabeled data with label
  propagation}.
\newblock  (\bibinfo{year}{2002}).
\newblock


\end{thebibliography}

\end{document}